# The missing ingredient in effective-medium theories: Standard deviations

Craig F. Bohren[1,*], Xuerong Xiao[2], and Akhlesh Lakhtakia[2]

[1]Department of Meteorology, Pennsylvania State University, University Park, PA 16802, USA
[2]Department of Engineering Science and Mechanics, Pennsylvania State University, University Park, PA 16802, USA

**Abstract**: Effective-medium theories for electromagnetic constitutive parameters of particulate composite materials are theories of averages. Standard deviations are absent because of the lack of rigorous theories. But ensemble averages and standard deviations can be calculated from a rigorous theory of reflection by planar multilayers. Average reflectivities at all angles of incidence and two orthogonal polarization states for a multilayer composed of two kinds of electrically thin layers agree well with reflectivities for a single layer with the same overall thickness and a volume-weighted average of the relative permittivities of these two components. But the relative standard deviation can be appreciable depending on the angle of incidence and the polarization state of the incident illumination, and increases with increasing difference between the constitutive parameters of the two layers. This suggests that average constitutive parameters obtained from effective-medium theories do not have uniform validity for all calculations in which they might be used.

Electromagnetic constitutive parameters such as permittivity and permeability are average response functions. For a pure molecular or atomic material such as water or glass or gold, the averaging volume is of order the cube of the wavelength of an exciting electromagnetic wave. At wavelengths well into the ultraviolet, the number of molecules per cubic wavelength is so large, even for gases, that these averages are usually sufficient for describing reflection and refraction because of optically smooth interfaces for any angle of incidence and polarization state, reflection and transmission by thin films and multilayers, and scattering and absorption by particles of any size, shape, and composition. Because there are so many molecules in the averaging volume, the standard deviation of the response function relative to its average is often negligibly small.

To an electromagnetic wave, a particle much smaller than the wavelength in both the material of the particle and the surrounding material (electrically small) is no different from a giant molecule with a very large polarizability. Thus a composite material consisting of, say, many small particles suspended in a continuous matrix (e.g., colloidal gold in aqueous suspension) should be characterized to good approximation by an average or effective permittivity (we take the permeability to be that of free space). But because the number density of particles in a composite material is much less than molecular number densities, we expect standard deviations for such a material to be appreciably greater than those for a molecular material. By particle we mean a bound aggregation of sufficiently many atoms or molecules that it can be assigned macroscopic properties such as temperature, pressure, density, and permittivity.

Effective-medium theories for composite materials have a long history, with contributions from Poisson, Mossotti, Clausius, Maxwell, Rayleigh, Maxwell Garnett,

Bruggeman, and others. For a compendium of classic papers on effective-medium theories as well as more modern papers, see Ref. [1].

It is often implicitly assumed that effective-medium theories apply to randomly inhomogeneous materials (as opposed to, say, an array of identical spheres at sites on a regular lattice). But a random material is not a single material, rather the name of an ensemble of many systems with the same volume fraction of particles of given composition suspended in a given material, distributed randomly in space and possibly in size, shape, and (if non-spherical) orientation.

Because an effective-medium theory yields only averages, two such theories for the same randomly inhomogeneous material cannot be legitimately compared, or a particular theory compared with measurements, without knowing standard deviations. And there's the rub. To our knowledge, standard deviations for composite random materials have not been calculated, and for good reason: lack of a rigorous theory of such materials.

Faced with this lack, to obtain some insights we turn to a composite system for which a rigorous theory does exist: a multilayer. Reflection and transmission by any number of layers can be calculated using the matrix method [2]. For example, for a normally incident electromagnetic wave ($E_i, H_i$), the electromagnetic wave reflected ($E_r, H_r$) and transmitted ($E_t, H_t$) by $N$ layers is

$$\begin{pmatrix} E_i + E_r \\ H_i - H_r \end{pmatrix} = \mathbf{M}_1 \mathbf{M}_2 ... \mathbf{M}_N \begin{pmatrix} E_t \\ H_t \end{pmatrix}, \quad (1)$$

where the characteristic matrix of a layer of thickness $d$, with wavenumber $k$ and intrinsic impedance $Z$ is

$$\mathbf{M} = \begin{pmatrix} \cos kd & -iZ \sin kd \\ -i \sin kd / Z & \cos kd \end{pmatrix}. \quad (2)$$

(Similar expressions hold for two orthogonal incident waves at arbitrary incidence.) If $2\pi |n_j| d_j / \lambda = 1$, $\mathbf{M}_j$ is approximately

$$\begin{pmatrix} 1 & -i2\pi d_j Z_0 / \lambda \\ -i2\pi (\varepsilon_j / \varepsilon_0) d_j / \lambda Z_0 & 1 \end{pmatrix}, \qquad (3)$$

where $\lambda$ is the free-space wavelength, $d_j$ is the thickness of the $j$th layer with permittivity $\varepsilon_j$ and refractive index $n_j = \sqrt{\varepsilon_j / \varepsilon_0}$, $\varepsilon_0$ is the permittivity of free space, and $Z_0$ is the impedance of free space. Harmonic time-dependence $\exp(-i\omega t)$ with circular frequency $\omega$ is assumed, and the permeability of all layers is that of free space $\mu_0$. If quadratic and higher powers of $2\pi |n_j| d_j / \lambda$ are neglected, the matrix of a multilayer with total thickness $h = \sum_j d_j$ is independent of the order of the layers and approximately,

$$\begin{pmatrix} 1 & -i Z_0 2\pi h / \lambda \\ -i2\pi (\varepsilon_{av} / \varepsilon_0) h / \lambda Z_0 & 1 \end{pmatrix}, \qquad (4)$$

where

$$\varepsilon_{av} = \sum_j \varepsilon_j f_j \qquad (5)$$

and $f_j = d_j / h$. $\varepsilon_{av}$ is a weighted average depending on only the volume fractions and permittivities of the layers. Of the geometrical properties of the multilayer, only the total thickness (relative to $h$) of all layers with the same composition, not their individual thicknesses, determines $\varepsilon_{av}$. For a two-component multilayer

$$\varepsilon_{av} = f \varepsilon_a + (1 - f) \varepsilon_b, \qquad (6)$$

where $f$ is the volume fraction of the component with permittivity $\varepsilon_a$. This equation, subject to assumptions underlying its derivation, is valid for all angles of incidence and both linear polarization states of the incident illumination provided that the permeabilities of the layers are equal and the ratio $\varepsilon_a / \varepsilon_b$ is neither too large nor too small.

Equation (6) is correct in the limits $f \to 0$ and $f \to 1$, which suggests that it is likely to be most accurate for $f = 1$ and $f \approx 1$, least accurate for intermediate values, say, $0.3 < f < 0.7$. The average Eq. (6) is an analytical expression. Although no such expression exists for the standard deviation, we can compute it as follows:

The reflection coefficient $\tilde{r}$ and transmission coefficient $\tilde{t}$ for a plane wave incident at angle $\theta$ on a two-component $N$-layer in free space can be calculated using

$$\begin{pmatrix} M_{11}\cos\theta - M_{12}/Z_0 & \cos\theta \\ M_{21}\cos\theta - M_{22}/Z_0 & 1/Z_0 \end{pmatrix} \begin{pmatrix} \tilde{r}_p \\ \tilde{t}_p \end{pmatrix} = \begin{pmatrix} M_{11}\cos\theta + M_{12}/Z_0 \\ M_{21}\cos\theta + M_{22}/Z_0 \end{pmatrix} \tag{7}$$

for $p$-polarization and

$$\begin{pmatrix} M_{11} + M_{12}\cos\theta/Z_0 & -1 \\ M_{21} + M_{22}\cos\theta/Z_0 & \cos\theta/Z_0 \end{pmatrix} \begin{pmatrix} \tilde{r}_s \\ \tilde{t}_s \end{pmatrix} = \begin{pmatrix} -M_{11} + M_{12}\cos\theta/Z_0 \\ -M_{21} + M_{22}\cos\theta/Z_0 \end{pmatrix} \tag{8}$$

for $s$-polarization. The matrix elements $M_{ij}$ are obtained from

$$\begin{pmatrix} M_{11} & M_{12} \\ M_{21} & M_{22} \end{pmatrix} = \exp(i\mathbf{A}_N d_N)\exp(i\mathbf{A}_{N-1} d_{N-1})...\exp(i\mathbf{A}_1 d_1), \tag{9}$$

where

$$\mathbf{A}_j = \begin{pmatrix} 0 & \omega\mu_0\{1-(\varepsilon_0/\varepsilon_j)\sin^2\theta\} \\ \omega\varepsilon_j & 0 \end{pmatrix} \tag{10}$$

for $p$-polarization, and

$$\mathbf{A}_j = \begin{pmatrix} 0 & -\omega\mu_0 \\ \omega\varepsilon_0 \sin^2\theta - \omega\varepsilon_j & 0 \end{pmatrix}. \tag{11}$$

for *s*-polarization. Either $\varepsilon_j = \varepsilon_a$ or $\varepsilon_j = \varepsilon_b$. The total number of layers is $N = N_a + N_b$, where $N_a$ is the number with permittivity $\varepsilon_a$ and thickness $d_a$, and $N_b$ is the number with permittivity $\varepsilon_b$ and thickness $d_b$. The volume fraction $f$ of the a-component is

$$f = \frac{N_a d_a}{N_a d_a + N_b d_b}. \tag{12}$$

If $d_a$ and $d_b$ are fixed, then for fixed $N$ and $N_a$, the total thickness of the multilayer is fixed. But the order of the layers is variable, and two (or more) layers of the same component material can be adjacent to each other. This corresponds to clumping of particles in a particulate composite material, which often is difficult to eliminate completely. To ensure that each layer is electrically thin we take

$$d_j = 0.1\lambda / 2\pi |n_j| \quad (j = a, b). \tag{13}$$

We generate an *N*-element array of permittivities, chosen randomly to be $\varepsilon_a$ or $\varepsilon_b$ subject to the constraint that $N_a$ is fixed. For each such array, and a fixed angle of incidence $\theta$, the reflectivity $R = |r|^2$ is calculated for the two polarization states from Eqs. (7)-(11). The average reflectivity $\langle R \rangle$ and its standard deviation are calculated for many such arrays and compared with $R(\varepsilon_{av})$, where $\varepsilon_{av}$ is the weighted average Eq. (6).

Figure 1 shows calculations for 250 arrays with $N = 50$, $n_a = 1.95$, $n_b = 1.4$, and $\lambda = 550$ nm. These refractive indices are for hypothetical materials with a permittivity ratio of about 2. $N_a = 30$ is chosen to give a volume fraction $f = 0.52$. For both incident polarization states $\langle R \rangle$ is approximately equal to $R(\varepsilon_{av})$ for all $\theta$. But the relative standard deviation is appreciable, as high as 20%. Perhaps more important, the relative standard deviation is not uniform, varying both with the angle of incidence and the polarization state. Calculations for higher and lower $f$ are similar.

Calculations also were done at $\lambda = 650$ nm for two layers composed of real materials, silicon dioxide ($SiO_2$) and cuprous oxide ($Cu_2O$). The refractive index of $SiO_2$ is 1.456 [3], and that of $Cu_2O$ is 2.90 + $i$0.1 [4]; hence, the ratio of their permittivities is about 4. Figure 2 shows calculations for multilayers with a cuprous oxide volume fraction 0.43. The maximum relative standard deviation, 40%, is even higher than in Fig. 1. Indirect evidence that these standard deviations are realistic is measurements of 90°scattering by single evaporating glycerol droplets (diameter $d \approx 6\,\mu m$ and $\approx 18\,\mu m$) containing about 1% by volume of polystyrene latex spheres ($d$ = 30-105 nm) [5]. A consequence of this inhomogeneity is fluctuations about the mean (up to 30%) of scattering as a function of droplet diameter, which increase with increasing latex sphere size.

What these simple calculations suggest, but do not prove, is that average permittivities of composite particulate materials may be accompanied by appreciable standard deviations. Also, such permittivities do not necessarily have the same unrestricted validity as those of molecular materials (which also are averages). Permittivities of molecular materials are often used without hesitation for calculating many different quantities. But our results suggest that the relative error in calculations using average permittivities of composite particulate materials can depend on what they are used for.

Reflectivity calculations using Eq (6) need not agree exactly or nearly so with calculations of the mean. If the former calculations lie within a standard deviation of the mean they can be said to agree with reflectivities for a random medium. It is easy to lose sight of this because we are accustomed to looking at two curves and saying that they agree or disagree. But for statistical calculations exact agreement doesn't exist. All calculations lying within an interval are equally correct. And the same is true of measurements made on only one member of a large ensemble of similar samples. We cannot know how close an individual measurement is to the mean of a large number of similar measurements unless they are made, which they often are not.

.__________________

*bohren@meteo.psu.edu

Figure Captions

1. Reflectivity as a function of angle of incidence for a two-component 50-layer multilayer averaged over 250 random arrays. (a) Incident *p*-polarization. (b) Incident *s*-polarization. Dashed lines show the average $\langle R \rangle$ plus or minus one standard deviation. The volume fraction of the component with refractive index $n_a = 1.95$ is 0.52; $n_b = 1.4$ and the free-space wavelength is 550 nm.

2. Reflectivity at 650 nm as a function of angle incidence for a two-component 50-layer multilayer composed of cuprous oxide (volume fraction 0.43) and silicon dioxide. (a) Incident *p*-polarization. (b) Incident *s*-polarization. Dashed lines show the average $\langle R \rangle$ plus or minus one standard deviation.

Figure 1

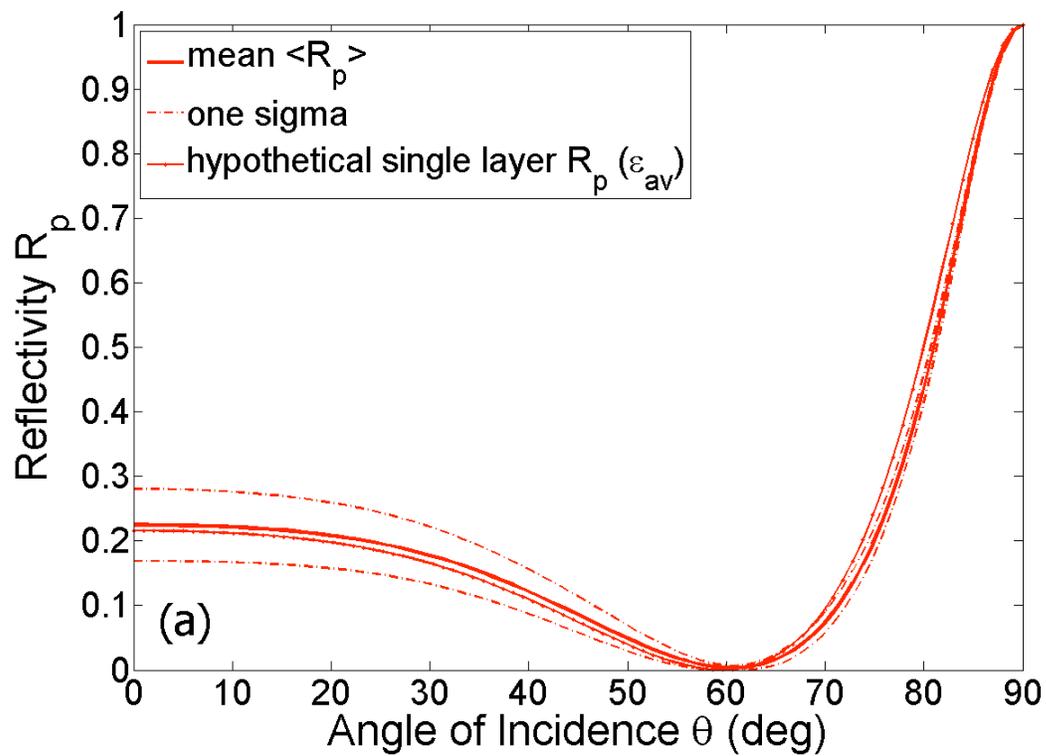

(a)

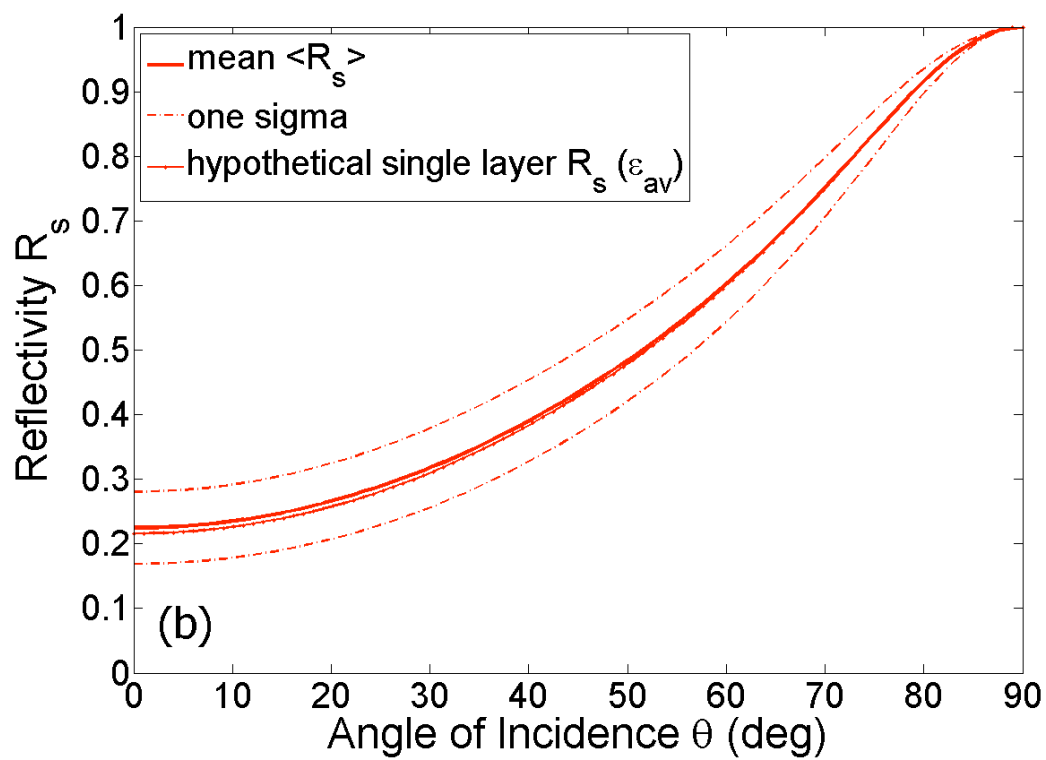

(b)

Figure 2

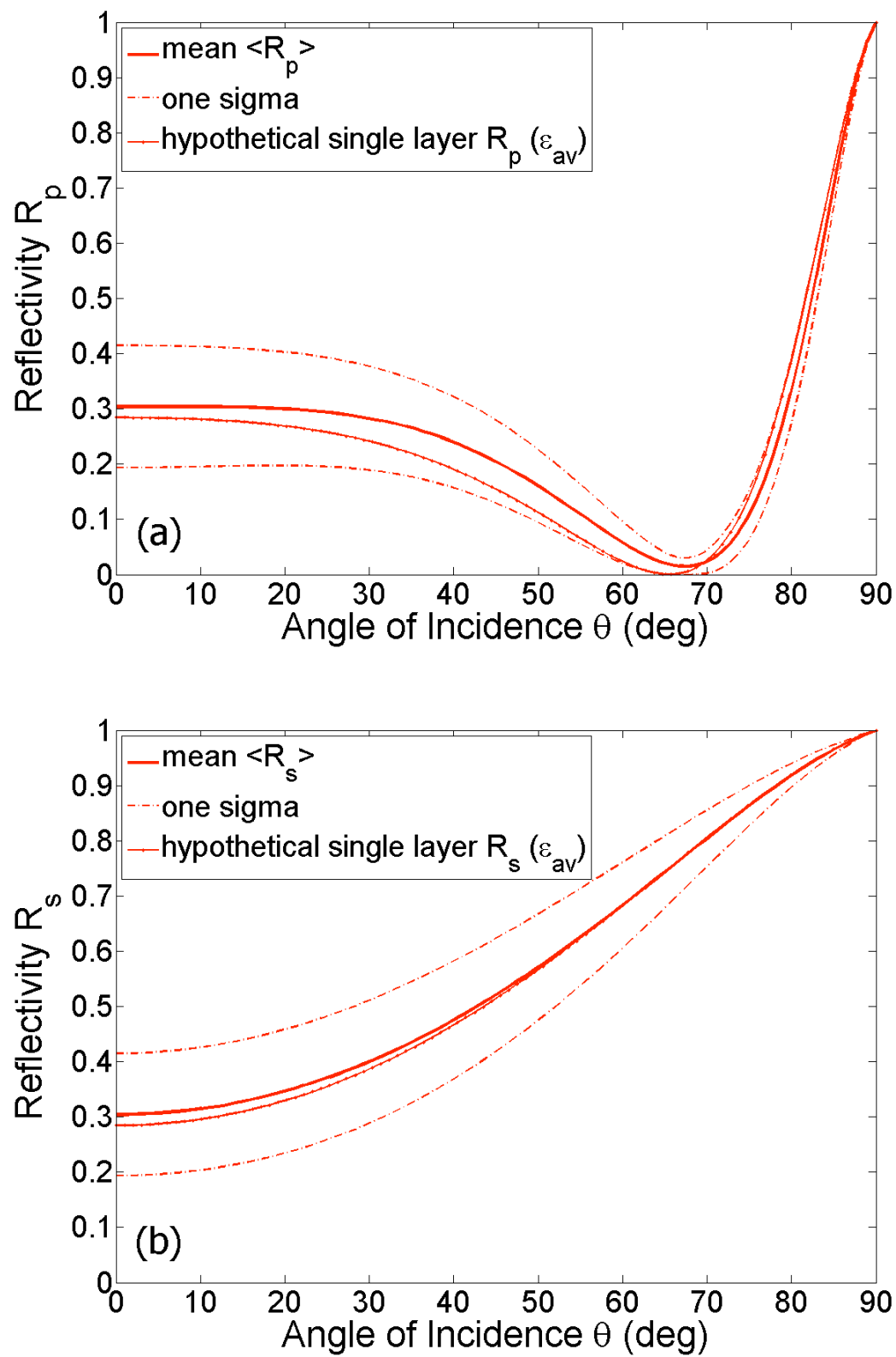